\documentclass[12pt]{article}
\title{           
Relativistic two-boson  system  in presence of electromagnetic  plane  wave}

\author{Ph. Droz-Vincent\\[2mm] {\em LUTH  \      Observatoire de Meudon}~\footnote{
 Observatoire de Paris-Meudon, Universit\' e Paris-Diderot,  5 place Jules Janssen 92195 Meudon, France }
}
 \date{  \          }
\newcommand {\ttt}{{ \scriptscriptstyle T}} 
\newcommand {\aaa}{{ \scriptscriptstyle A}}
\newcommand {\BBB}{{ \scriptscriptstyle B}}
\newcommand {\lll}{{\scriptscriptstyle L}}
\newcommand {\onerom}{ { \scriptscriptstyle {\rm I} } }
\newcommand {\tworom}{ {\scriptscriptstyle  {\rm II}} }
\newcommand  {\eeq}{\end{equation}}

\newcommand {\del}{\delta}
\newcommand{\soulP}{\underline P}

\newcommand{\soulT}{\underline T}
\newcommand{\soulG}{\underline  G}

\newcommand{\soulVprim}{\underline  V'}
 \newcommand {\dron}{\partial}
\newcommand  {\beq}{\begin{equation} }
\newcommand  \half {  {1 \over 2} }

\newcommand{\Zhat}{\widehat Z}
\newcommand{\soulZhat}{\underline {\widehat Z}}
\newcommand  {\noi}{\noindent}
\newcommand  {\doot}{\!  \cdot  \!  }
\newcommand  {\disp}{\displaystyle}
\newcommand  {\zer}{ {(0)} }
\newcommand {\sig}{   \sigma   }

\newcommand {\vareps}{\varepsilon}
\newcommand {\alp}{\alpha}
\newcommand {\lam}{\lambda}

\newcommand{\gam}{\gamma}
\newcommand{\ome}{\omega}  
\newcommand{\cale}{{\cal  E}}  
\newcommand{\calf}{{\cal  F}}

\newcommand{\soulK}{\underline K}

\newcommand{\batonR}{{\mbox{I\hspace*{-1mm}R}}}

 \newtheorem{prop}{Proposition}
 \newtheorem{theo}{Theorem}
\newcommand{\beprop}{\begin{prop}}
\newcommand{\betheo}{\begin{theo}}
\newcommand{\enprop}{\end{prop}}
\newcommand{\entheo}{\end{theo}}
\begin{document}
\maketitle     \abstract{The relativistic two-body problem is considered for spinless  particles  subject to  an external 
 electromagnetic field. When this  field is  made of the  monochromatic superposition of two  counter-propagating
 plane waves (and provided the mutual interaction between particles is known),  it is possible to write down  explicitly   
a pair of  coupled wave equations  (corresponding to a pair of   mass-shell constraints)
  which   takes into account    also  the field contribution. These equations are manifestly covariant; constants of the motion  are  exhibited,   so   one ends up with   a   reduced  problem
 involving  five degrees of freedom.}

\section{Introduction}
\noi   

A large litterature is devoted to the effect of applying a laser field to various objects  (electron beams,    two-level atoms, etc).  In early papers [1-3]
  the laser wave  was  modelled as purely electric,  the effect of a  magnetic contribution being introduced more recently~[4-5].
  Experimental devices are concerned with a wide variety of atoms,  but theoretical developments so far have mainly    considered  an atom as a two-level system; however  this  simplification (which  seems to be adequate in most  realistic cases)       does not fully take into    account the few-body structure of atoms.    
So, at least in principle,  a  more accurate  treatment is desirable;  as a  very first step, it is natural to consider a two-body problem involving two  charges  in interaction  in the presence of a plane  electromagnetic wave. 
In this spirit   hydrogen-like bound states are   the most simple  targets  offering   the two-body structure,   
but for simplicity we focus here on spinless particles.

A  relativistic  two-particle system  can be covariantly  described  by  a pair of  coupled Klein-Gordon  equations referred to as  mass-shell constraints;
these constraints determine the evolution of a wave function which depends on a pair of  four-dimensional arguments~[6-16].
This approach is suited for  situations where particle creation can be neglected whereas other relativistic effects must be taken into account.
In this  article   we   consider  a  pair of  charged  particles undergoing  some mutual interaction whereas  they are  coupled with an external   electromagnetic field. 
The following assumptions are made:

\noi The external  field is not affected by the motion of the system, 

\noi This external field does not create pairs.  

\noi We neglect  the radiation emitted by the charges in their motion.

\noi     Thus the  validity of this  picture is implicitly limited  by some conditions involving the strenght and the analytic shape of the field,  the  Compton wavelength  of the  particles,  etc.
The conditions  under which    laser fields {\em  may} create  pairs have been widely discussed in the litterature (see Ref. ~[4,17]
 for fermionic pairs), but  here we are interested
 in the cases where pairs {\em cannot} be created.

In the absence of external field the contact of the constraint formalism with QED is well established,  through the quasi-potential approach~[18-21]
   or by reducing  the Bethe-Salpeter  (BS) equation~[22,23],
 and  realistic mass-shell constraints can be exhibited, see  Crater {\em et al.}~\cite{crater}, see  Sazdjian and Jallouli~\cite{jallou97}  when the mutual interaction also is of electromagnetic nature. 

\noi  In the presence of external field, it seems   natural  to perform  in the same manner   a reduction of the BS equation and  so derive the coupled wave equations.
This task was carried out by Bijtebier and Broekaert  \cite{bijbroek} in the case of {\em static} external potentials, but the tractability of the method for  time-depending cases is problematic.
Moreover it is sometimes interesting  to consider also  phenomenological models of (mutual) interaction,  not provided with  a  given  BS kernel.

\noi So we prefer a  direct coupling to the field, constructed with  help of  these  reasonable prescriptions: correct limits when one of the interactions vanishes,   and  invariance under
 the spacetime  symmetries   that survive application of the field. 

\medskip
In the present  paper  the explicit form  of  the mass-shell constraints is supposed to be known for the isolated system.  Naturally the  mutual interaction between the charges  may be 
simply   electromagnetic,  but in several  models  this  electromagnetic term is neglected in front of other forces of phenomenological origin. 

When an external field is applied, we face this old  problem  that relativistic interactions cannot be just linearly composed; this was soon understood by Sokolov  in   early attempts
 to construct n-body systems, in the  context  of a very different approach~[27,28].
 In our framework,  the trouble  is that    the coupled wave equations must remain  compatible one with the other, and this requirement is essentially  nonlinear.
 As a result  we   generally  do not know  how   to write down in closed form a new pair of compatible  wave  equations     that takes  into account also the coupling to the external field. 

 Nevertheless this problem becomes solvable when the external potential corresponding to  the external field has a particular symmetry  that we proposed to call 
 {\em strong} translation invariance.  
 The solution is furnished by an {\sl  ansatz}  elaborated in such a way that  wave equations have the correct limits when either the mutual interaction or the external potential is turned off; these equations also 
have  the correct nonrelativistic limit.    Moreover in this context, under reasonably  general  assumptions,  the principle of isometric invariance~\cite{IJTP2009} is satisfied, say:
 the coupling to the external field {\em must  respect  the surviving isometries};  in other words   the spacetime  isometries that would survive (as symmetries of the system)
 application of the external field to  mutually  independent   particles   should   remain   symmetries of the  system  when these  particles  also undergo  the  mutual interaction.
This property  was not considered when the ansatz was proposed for the first time; having it  finally satisfied is a bonus.

   Our  method    was  initiated by J. Bijtebier~\cite{bij89} in the  simplest  case of  a  stationary  external  interaction.   Introducing  the notion of {\em strong} translation invariance,  we carried out a 
systematic generalization~\cite{drozNCim}    which can be applied to a lot of more complicated  situations.  
  
Focusing on the case where external interaction is due to   some  electromagnetic plane  waves,  we already  started  many years ago a discussion about the tractability of the ansatz~[32,33]
 and we pointed out the interest of considering a monochromatic superposition of two waves; but the subject was not developped further. 
In the present  paper our purpose is to  construct  (for the first time explicitly)  the mass-shell constraints  and  to exhibit the first integrals of the motion, allowing for a reduction of the number of degrees of freedom.
By the same token several  complicated calculations and involved arguments given  in our  aforementioned articles will be replaced by a more comprehensive exposition.

\medskip

Section 2  is devoted to a  survey   of the general  formalism we employ, in both  the one-body and two-body cases.   

\noi In Section 3 we return to   one-body   motion, focusing on the case where external field is made of  plane electromagnetic waves.    Indeed the  one-body motion must be  considered first, because   strong translation invariance,  which helps   us to formulate   the {\em  two}-body  problem in closed  form, is basically  defined as  a  property of the {\em one}-body motion in the field.  We present  and discuss   two  cases of external fields:

\noi   i) the  single monochromatic plane wave, 

\noi       ii)  a  monochromatic superposition of two such waves, which can be seen as  made of two counter-propagating  waves.

\noi Here we  show  how  strong translation invariance arises    and  we discuss its  usefulness with repect to our purpose.
 
\noi Then,  in Section 4,  the details of   formulating the  two-body wave equations  in presence of the monochromatic superposition   are displayed.
Section 5 is devoted to concluding remarks.

\medskip
{\sl Notation, terminology} $ \    $  Signature $+---$, units  $c=\hbar =1$.
{\em  Isometry}  refers to any transformation of the Poincar\'e  group acting on spacetime. 
Let us call {\em surviving symmetries}  of a system those symmetries that are not destroyed by application of the external field. 
Particle labels $a,b$ run from 1 to 2.

\section{Basic formulas}
 This  section is not limited to the  case of electromagnetic plane waves; here we intend to consider  a   more   general case where the external interaction enjoys  the appropriate symmetry.
 
\noi
In fact strong translation invariance which (when it arises)  helps us to formulate the two-body  problem, is primarily    a  property of the one-body motion in the field.
 Therefore in the next  subsection  we first consider a {\em single} test particle submitted to an external field. 

\subsection{The one-body wave equation,  symmetries  }
Let $x, p$ be the canonical variables  satisfying the standard commutation relations    $ [x^\alp , p_\mu  ] = i \del ^\alp _\mu  $.
  The one-body wave function is a scalar on spacetime, say  $\psi(x)$. The  Klein-Gordon equation $ 2K  \psi  =   m^2  \psi    $    is written using the {\em  half-squared mass operator} 
\beq  K = \half  p^2 +  G ,    \label{1K-G}         \eeq
 where the interaction term $G$ will be called  {\em external  interaction potential}. It is a {\em scalar}, not to be confused with the   electromagnetic  { \em  vector}  potential $A_\mu$.

 This scheme may be obtained by quantization of a classical relativistic (and manifestly covariant) formalism where the equations of motion stem from  a {\em scalar} generator  
$ K_{\rm cl} =  \half  p_{\rm cl}  \cdot  p _{\rm cl}  +  G_{\rm cl} $,   the  {\em eight}  canonical  variables     $x _{\rm cl} ^\alp , { p _{\rm cl}} _\alp $  being  conjugate in terms of their 
   Poisson brackets;  the generator is an obvious constant of the motion and is interpreted as  half of the squared mass.   Although this approach is not very popular   among physicists,
  it is very  simple and natural from    a geometrical  point of view, as it rests on the symplectic structure of the eight-dimensional  phase space.
In this formulation the evolution parameter is affine, that is proportional to the proper time;    
  the role of the  half-squared mass with respect to this parameter  is  analogous  to the role of the energy  with respect to the absolute time in  Newtonian mechanics.

\bigskip
\noi  {\sl Strong translation invariance}. 

\noi  Only some fields give rise to strong translation invariance. The most simple example corresponds to an interaction potential of the form 
   $G= G(\vec { x }, \vec { p})$, with obvious notation using  the lab frame, supposed to be unique\cite{bij89}.   

\medskip 
 \noi   {\em  Definition $\       $  Any phase-space function is  {\em  strongly} translation invariant along direction   $w$ when it  commutes with both     $w \cdot x $     and     $w \cdot p $}.
  
\noi When it commutes with  {\em only}    $w \cdot x $   it is  said {\em  simply}   translation invariant.      
    
\medskip
\noi  We are interested in  having the external potential strongly translation invariant.
  In the space of four-vectors, the directions leaving  $G$  strongly  translation invariant, when they exist, span a linear space $E_L$ called {\sl longitudinal space}.
For our purpose it is essential that   $E_L$ admit an orthonormal basis~\footnote{
   But some situations of physical interest are not included in this case: for instance when the external field is a {\em   single}  monochromatic  plane wave, it turns out that the wave vector 
 is a direction of strong translation invariance; but it   is a null vector, so  in this case  $E_L$ admits no orthonormal frame.};
 in this situation   referred to as  {\em normal}, the space of four-vectors  is an orthogonal direct sum  $ E  = E_L    \oplus E_T $ where $E_T $ is the  {\sl transverse  space}. 

\noi  Any  four-vector  $\xi$  can be written as  
$$\xi = \xi_\lll   +   \xi_\ttt,   \qquad    \quad     \xi_\lll   =  \tau   \xi   $$
with   $\tau $   projector  onto   $E_L$.   The  canonical variables are  split accordingly,  say
$x = x_\lll   +   x_\ttt  ,  \qquad             p =p_\lll   +  p_\ttt     $.  
 Phase-space functions are called  longitudinal (resp. transverse) when they depend only on the longitudinal (resp. transverse) canonical variables, equivalently they commute
 with all the transverse (resp. longitudinal) canonical variables.

\noi The case where $E_L$ fails to admit an orthonormal basis will be referred to  as   {\em degenerate}.

\beprop { The interaction potential is a transverse quantity}.  \enprop

\noi Proof $\    $Let  $w_\aaa$ be an orthonormal basis of $E_L$ ( the number of values taken by  indices  $\aaa , \BBB$  is just the dimension of  $E_L   $ ),   
 so $w_\aaa \cdot w_\aaa =  \vareps _ \aaa $, with   $\vareps _\aaa = \pm 1$ depending on the signature of $E_L$.
The projector onto $E_L$ is
$$ \tau ^\alp _\beta   =  \sum _\aaa   \vareps _\aaa  w_\aaa  ^\alp   w _{\aaa  \beta}    $$
$$ x_\lll ^\alp =         \tau ^\alp _\beta   x^\beta  =        \sum _\aaa   \vareps _\aaa  w_\aaa  ^\alp (w_\aaa \cdot x)      $$
\beq   [G , x_\lll ^\alp ]  =          \sum _\aaa   \vareps _\aaa  w_\aaa  ^\alp   [G ,  w \cdot x  ]   = 0   \eeq
Similarly   $ p_\lll ^\alp =         \tau ^\alp _\beta   p^\beta $  and one finds 
 \beq     [G , p_\lll ^\alp ]  =  0              \eeq
[]

\subsection{Two-body wave equations}

\medskip
   The wave function is  $\Psi (q_1 , q_2)$ where  the points  $q_1 , q_2$ in spacetime   are canonically    conjugate  to the momenta
  $\disp  p_a ^\alp  =-i  {\dron  \over \dron q_a ^\alp}$. 
Noice that the arguments of $\Psi$  are  denoted as  $q_1 , q_2$ rather than    $x_1 , x_2$ because their classical  ( {\em  i.e.}  non-quantum ) analogs 
 may fail to  coincide with  the positions in the whole phase space of classical relativistic dynamics~[34-36].

The coupled wave equations involve the half-squared-mass operators $H_a$ as follows
 \beq  2  H_a  \     \Psi  =  m_a ^2   \   \Psi  ,   \qquad   \        a=1, 2        \label{weq2}   \eeq 
where the commutator   $ [H_1 , H_2 ]  $ must vanish. First integrals   (also called constants of the motion) are characterized by 
commuting  with {\em both} $H_1$  and  $ H_2  $. 

\noi The wave equations      (\ref{weq2}) can obviously be replaced by their sum and difference. 
Moreover  it is convenient to define
$$ P  = p_1  + p_2 ,   \qquad  Q = \half ( q_1 + q_2) ,     \qquad             y = \half  (p_1 -p_2 ) ,  \qquad z= q_1  - q_2  $$
and  $  j_a = q_a \wedge  p_a     $.
The Lie algebra of the  Poincar\' e group  is spanned by its  generators   $P^\alp$  and    $J_{\mu \nu}$, where  $J =Q \wedge P +  z \wedge y   =  j_{1 }  +    j_{2 }  $.

\noi In the absence of {\em  mutual}   interaction we would simply  have the following  equations
\beq  2  K_a  \     \Psi  =  m_a ^2   \   \Psi                    \eeq  
\beq   K_a  =    \half  p_a ^2  +   G_a          \label{defKa}  \eeq
where $G_a (q_a , p_a )$ is the {\em external}  interaction potential  acting on particle $a$; in an external electromagnetic field,
$K_a$ and $G_a$ are obtained  by replacing  $x,  p$ and the charge $e$  respectively by $q_a , p_a$ and  $e_a$  in   $K$ and in $G$, see \cite{IJTP2009}.

\medskip
\noi  {\sl Definition}$\qquad \    $  The couple of potentials  $G_1 , G_2$ is {\em strongly invariant  along direction $w$}  when each potential separately is
 strongly invariant along $w$ in the one-body sector,  say 
$$  [ G_1  , \     w \cdot q_1 ]   =  [ G_1  , \      w \cdot p_1  ]   =     [ G_2  , \      w \cdot q_2 ]   =        [    G_2  , \     w \cdot p_2 ]           =0        $$        
Again  the {\sl  longitudinal space}  $E_\lll$  is defined as  the span of strong translation  invariance directions, and the question arises as to know whether  $E_\lll$ admits an orthonormal basis.
In the {\em normal case} (that is when it does admit such a basis)  Proposition~1 entails that $G_1$ and $G_2$ are {\em  transverse}.

\medskip       
If the {\em external}  field were zero  , we would  have (the superscript $\zer$ refers to   the absence of external field) 
\beq    2 H _a^\zer  \    \Psi  =  m_a ^2    \   \Psi                          \eeq
where  we could have     
$  \disp   H _a^\zer   =   \half  p_a ^2  +    V^\zer _a     $ but  in this  work  we assume a {\em unipotential model}, say $V^\zer _1     =    V^\zer _2  =     V^\zer $, hence
\beq    H _a^\zer   =   \half  p_a ^2  +    V^\zer        \label{Vzerunipot}                  \eeq
and we  suppose that  mutual interaction is explicitly  known   when   the external field vanishes, in other words  $V^\zer $ is given.

For simplicity let us assume that  the mutual interaction takes on the form    
\beq    V^\zer =  f   (Z,  P^2 ,   y \cdot  P  )      \label{defVzer}    \eeq 
where   
\beq   Z = {z \cdot z} \     P^2  -  (z \cdot P )^2       \label{defZ}      \eeq 
is the main ingredient;  so  the three dynamical variables in $V^\zer$ are mutually commuting.

\noi Let us  stress that the simplification which results from assuming     (\ref{Vzerunipot})(\ref{defVzer})  leaves   still enough generality  to encompass
 a wide class of mutual interactions;  as an example see the model of electromagnetic interaction derived from Feynman's diagrams by Jallouli and Sazdjian~\cite{jallou97}.

\bigskip
\noi When all interactions are present,  eqs.(\ref{weq2})  hold with 
\beq         H _a =  K_a   +  V           \label{unipotext}   \eeq
where  $V$ is a suitable modification of  $V^\zer$, to be constructed in order to 
satisfy the vanishing of      $   [ H_1 ,   H_2  ]$   and to  reproduce the correct  limits when either of the 
interactions is absent.     

\noi  In  general the compatibility condition
$$    [K_1 - K_2 ,   V ] =  0    $$
cannot be solved for $V$ in closed form.  But when the external potentials $G_a$ enjoy  some special symmetry, this equation can  be transformed  to a tractable problem,
 owing to a change of representation,  as follows

\noi Let $ {\cal O}$ be any operator;  
  the {\sl external-field representation} is formally defined by
\beq    \Psi '  =  {\rm e}  ^{i B}  \      \Psi  ,  \qquad \       {\cal  O}' =  {\rm e}  ^{i B}  {\cal  O}     {\rm e}  ^{-i B}        \label{transbij}   \eeq
        where  $B$  suitably depends on the external potentials.  
Of course   we are left with the task of solving for  $V'$ 
\beq    [K'_1 - K'_2 ,  V']  =0             \eeq
which will be possible provided  $B$ is choosen such that $K'_1 - K'_2$ is computable and takes on a simple form.
Moreover  we have to  write down explicitly the  transformed of  system  (\ref{weq2}), which requires  that  both  $K'_1$  and  $K'_2$  be computable by  
(\ref{transbij}).

 The concept of {\em strong translation invariance, naturally extended to  the two-body sector,  offers a possibility to  carry out this program}, choosing $B$ such that 
\beq        K'_1 - K'_2  =  y_\lll \cdot  P_\lll              \label{difKprim}     \eeq

\bigskip

\noi  Before going further  a few remarks are in order.  

\noi In many  cases of interest, the presence of external potentials   does not   kill all the   isometries  of spacetime.  

{Example  $\     $ We shall see later on in Section~4  that, in an  electromagnetic field satisfying the  equations    (\ref{defAmu})(\ref{defW}) below,         
the survivors of the Poincar\'e algebra are, in an adapted frame
  \beq           P_1 = p_{ (1) 1 } +   p_{ (2) 1}   ,  
        \qquad    \qquad                P_2 = p_{ (1) 2 } +   p_{ (2) 2}        \label{alg2}        \eeq  
(in these   formulas  and whenever necessary in the sequel,  we put  {\em parenthesis around the  particle indices}).

\noi  In the above example $P_{\lll  1} =   P_1$ is the only  nonvanishing component of $P_\lll$. In contrast  $P_2$ is just another conserved quantity.

\medskip

\noi  {\sl Notation}  $\     $   In order to avoid confusing the square of a vector with its second contravariant component, we make the following convention:

\noi   use  covariant components for the momenta $P_\alp , y_\beta$, and contravariant  components for the coordinates $Q^\mu ,  z^\nu$.
So  $P^2$ stands for  $P \cdot P$,   but    $z^2$ is the second component of $z$,     whereas the square of $z$ is  explicitly noted as  
$z \cdot z$.  This convention also  holds  with  longitudinal and transverse parts, for example  $P_\lll^2 = P_\lll \cdot P_\lll$, etc,
  but  we write  the square of $z_\ttt$ as  $z_\ttt \cdot  z_\ttt$,  etc.

\medskip

In the remaining part of this section  we  are concerned  with  strong invariance in  the {\em  normal  case}   where (by definition)  longitudinal and transverse parts are well defined, 
 which corresponds to the existence of  an orthonormal basis in $E_L$ .
Under this assumption  it was proved that {\em  the longitudinal piece   of linear  momentum, say    $P_{\lll \alp}$,   is conserved}  (see  Section 3.2    of  \cite{drozNCim}).

\bigskip     
  
Formula    (\ref{difKprim}) is ensured by     taking            $B = T L$, 
where  $T$   and $L$       are respectively  transverse     and  longitudinal   operators suitably chosen, namely
 \beq   T =  y_\ttt  \cdot P_\ttt   +  G_1  - G_2               \label{defT}          \eeq
 \beq   L  =   {P_\lll   \cdot  z_\lll   \over     P_\lll   ^2 }      \label{defL}    \eeq
Note that only $T$  depends  on the external field. But the  denominator  in  $L$    requires some caution  in order to make sense; 
 so we are led to cut off  the space of states  a sector  corresponding to the  vanishing  eigenvalues   of  $P_\lll ^2$, as follows. 
The wave function can be considered as a function of     $Q$ and $z$, which can be   written as  a  Fourier expansion  with respect to  $ Q$,
\beq    \Psi  =  {1 \over (2 \pi )^ 2 }      \int    {\rm e} ^ {i K \cdot Q}  \        \Upsilon (K , z )  d^4  K                                \label{fouPsi}     \eeq
Introducing an arbitrarily small but positive constant $\epsilon$,   we shall  restrict  the support of  $\Upsilon  $ to
 the values of the vector $K_\alp$  satisfying    
  $$K_\lll  ^2   \geq  \epsilon$$
 Using once for all the Fourier development (\ref{fouPsi}) it is easy to check that the operators  $z, y, Q, P$ and  $1/P_L ^2$
respect the cut-off;  in other words,  any of these operators   maps  into itself  the space of  wave functions  whose Fourier transform satisfies the support condition above.

\noi  Naturally $\disp \     P_\alp = -i {\dron \over \dron Q^\alp} \    $  and  $  \     \disp        y_\alp = -i {\dron \over \dron z^\alp }    \      $    thus for instance we have that 
$$  { 1 \over P_L ^2}  \Psi =      {1 \over (2 \pi )^ 2 }      \int  {\rm e} ^ {i K \cdot Q} \    { 1 \over K_L ^2} \      \Upsilon (K , z )  d^4  K       $$ 
  $\    P_L$   is a constant of the motion and  we shall eventually focus on eigenstates of it; in that case the support condition will get trivially simplified.

\bigskip
Note that   $T$  
is manifestly transverse in (\ref{defT}); however   we  can write equivalently
\beq  T  =  K_1 -  K_2   -   y_\lll  \cdot  P_\lll              \label{vardefT}      \eeq
 
\noi    It is obvious in (\ref{defL}) that $P_\lll$ commutes with $L$, thus also with $B$, hence  $P'_\lll = P_\lll$. 

\noi  Similarly  transformation  (\ref{transbij})  brings no change in $L$, nor in $T$, say   $L' =  L , \         T' =T$ .

\noi The explicit form of  $K'_1$ and  $K'_2$ was  derived from  (\ref{defKa})         in  \cite{drozNCim}. 
In order to get rid of  a  clumsy  notation ($L^2 \not= L^\alp L_\alp$) we replace here the four-vector   $L^\alp$         proposed in  \cite{drozNCim}                           by its definition,
 say  $L^\alp = P_\lll ^\alp /  P_\lll \cdot  P_\lll$,  
 hence  for the sum
\beq  K'_1  +  K'_2 = K_1 + K_2  - 2T \,  {y_\lll \cdot  P_\lll \over    P_\lll ^2}  +  {T^2  \over  P_\lll ^2}      \label{somKprim}   \eeq   
and   (\ref{difKprim}) for the difference;                 
in fact the external-field representation was taylored for having that   (\ref{difKprim})  holds true.

\noi Finally, after defining
$$     \mu = \half (m_1 ^2  +  m_2 ^2 )  , \qquad  \         \nu   =  \half (m_1 ^2   -  m_2 ^2 )                   $$
the coupled wave equations   $H' _a  \Psi '  = \half m_a^2      \Psi '   $ obtained from (\ref{weq2})  take on this form
\beq   
  (K'_1 + K'_2  +  2V')  \Psi '  =    \mu    \Psi   '      \label{somprim} \eeq          
\beq  
    y_\lll \cdot  P_\lll    \Psi '  =   
\nu     \Psi   '        \label{difprim}    \eeq
(note that the latter equation does not depend on the  mutual interaction).

\medskip

The explicit form of $V'$ is constructed from that of  $V^\zer  $ as follows:
according to (\ref{difKprim}) we have to ensure that   
\beq         [ y_\lll \cdot  P_\lll , V' ]  = 0            \label{Vprimsolve}         \eeq
A solution is easy to find, introducing the no-field "limit" of  $B$, obtained by cancelling $G_1$  and  $G_2$ in $T$, say  
$$ B^\zer =  y_\ttt \doot  P_\ttt  \,       L       $$
indeed  we  observe that 
$$   {\rm  e}^{iB^\zer} \,  y   \doot  P  \,     {\rm e}^{-iB^\zer}  =  {\rm  e} ^{iB} \,     y   \doot  P \,     {\rm  e}^{-iB}    =     y_\lll \cdot  P_\lll      $$
and we know that  $Z$ commutes with   $y \cdot P$,  therefore  defining
$\Zhat$ as    the no-external-field limit of $Z'$, namely 
\beq  \Zhat  = {\rm  e}^{iB^\zer}   Z     {\rm e}^{-iB^\zer}         \label{formalZhat}   \eeq
it turns out that $\Zhat $  commutes with     $y_\lll \cdot  P_\lll $. 
\noi  Moreover  it is obvious that $P^2$    commutes with    $y_\lll \cdot  P_\lll $,  thus  finally  any function of   $\Zhat , P^2 , y_\lll  \cdot P_\lll$  
is  expected to solve    (\ref{Vprimsolve}). 

\medskip
\noi  The {\sl ansatz}  consists in constructing $V'$  from $V^\zer$ as follows
\beq  V ' = f ( \Zhat   ,  P^2 , y_\lll \cdot  P_\lll )     \label{ansatz}   \eeq    
where  $f$ is   the function given     in   (\ref{defVzer}).
It is not difficult to check that this formula yields the correct limits when either $f$ or  both $G_1$ and  $G_2$  vanish.

\noi Fortunately $\Zhat$ is  explicitly  computed; formula  (\ref{formalZhat})  yields

\beq  \Zhat = Z   + 2 (P_\lll ^2 \quad    z \cdot P -P^2 \quad    z_\lll \cdot  P_ \lll  ) L
+   P_\ttt ^2     P_\lll ^2        L^2       \label{defZhat}  \eeq
In \cite{IJTP2000}  this  formula  
 was  cast into this equivalent  but  more compact form 
\beq  \Zhat =  z_\ttt  \cdot  z_\ttt \        P^2   -  ( z_\ttt   \cdot P)^2   +  P^2  (z_\lll \cdot z_\lll   - { ( z_\lll  \cdot  P_\lll ) ^2 \over  P_\lll ^2})
 \label{7IJTP2000}    \eeq
 We emphasize that in Ref. \cite{IJTP2000}  all numbered formulas  from (1) up  to (7) included
are general and by no means limited to the case of constant electromagnetic field~\footnote{In contrast eq. (8) of that reference holds only
 if   $P_\lll$ is timelike, which is not the case eventually considered in the present paper.}.
Indeed inserting   (\ref{defL}) into  the middle term of  (\ref{defZhat}) we obtain
$$2 (P_\lll ^2 \   z \cdot P -P^2 \   z_\lll \cdot  P_ \lll  ) L   = 2  (z \cdot P)  (z_\lll  \cdot  P)   -   2{P^2 \over P_\lll ^2} (z_\lll  \cdot  P) ^2$$
$$2 (P_\lll ^2 \   z \cdot P -P^2 \   z_\lll \cdot  P_ \lll  ) L   = 2  (z \cdot P)  (z_\lll  \cdot  P)   -  2  (z_\lll  \cdot  P) ^2    (1+  P_\ttt ^2  / P_\lll ^2 ) , $$
splitting    $z \cdot P $ yields
$$2 (P_\lll ^2 \   z \cdot P -P^2 \   z_\lll \cdot  P_ \lll  ) L  =    2  (z_\lll  \cdot  P) ^2  +   2  (z_ \ttt \cdot P)  (z_\lll  \cdot  P)
    -    2  (z_\lll  \cdot  P) ^2 (1+  P_\ttt ^2  / P_\lll ^2 )    $$
Now  add  $Z$,  taking   (\ref{defZ}) into account; we find a  cancellation of    $(z_ \ttt \cdot P)  (z_\lll  \cdot  P)$ and  remain with
$$  Z  +      2 (P_\lll ^2 \   z \cdot P -P^2 \   z_\lll \cdot  P_ \lll  ) L   =      z \cdot z \      P^2  -  (z_ \ttt \cdot P)^2   -  (z_\lll  \cdot  P) ^2  (1 +  2   P_\ttt ^2  / P_\lll ^2 )$$
Owing to   (\ref{defZhat}),  $\Zhat$ is given by  adding  $ P_\ttt ^2    P_\lll ^2        L^2  $ to this quantity.
But  (\ref{defL}) implies that$ P_\ttt ^2  P_\lll ^2  L^2 =     (z_\lll  \cdot  P) ^2     P_\ttt^2  /  P_\lll  ^2$, thus
$$ \Zhat  =  z \cdot z  \      P^2  - (z_\ttt \cdot P )^2   -  (z_ \lll \cdot  P)^2  (1 +  P_\ttt^2  /  P_\lll  ^2 )         $$
$$ \Zhat  =  z \cdot z  \      P^2   - (z_\ttt \cdot P )^2   -  (z_ \lll \cdot P )^2    {P^2  \over  P_\lll ^2 }    $$
after splitting     $ z \cdot z \      P^2 $  we are left with       (\ref{7IJTP2000}),     
that is  formula (7) of   \cite{IJTP2000}. []

\noi  In order to develop (\ref{somprim}) we need to compute  the r.h.s. of (\ref{somKprim}).  

\noi Eq. (\ref{defKa})  implies     $  K_1 +K_2  = P^2 /  4 + y^2   + G_1  +  G_2 $      that we insert into  (\ref{somKprim}), hence 
\beq   K'_1 +K'_2  =  P^2 /  4 + y^2   + G_1  +  G_2    - 2T \,   {y_\lll \cdot  P_\lll \over    P_\lll ^2}  +  {T^2  \over  P_\lll ^2}     \label{varsomKprim}    \eeq

In the following sections we shall specify the external field.

\section{One-body motion in  electromagnetic waves}

 \noi
Consider first  the motion of a single charged  pointlike and  spinless body (treated as a   test particle)  subject  to {\em any}  electromagnetic field  $F =\dron \wedge A$.
Let  angular momentum  be  noted  as   $j = x \wedge p$.   In the  Klein-Gordon equation we have 
\beq 
 K = \half  (p- {e } A ) ^ 2 , \qquad \   
  G = - \half (e A \cdot p + e p \cdot A  -  e^2 A^2 )  
\label{defG}   \eeq

Note that
\beq    [  A^\mu , p_\nu  ] = i  \,   \dron _\nu A^\mu \   \label{Amupnu}      \eeq

\bigskip

\subsection{Single electromagnetic  plane wave} 

The behavior of a relativistic charged particle in an electromagnetic plane wave has been studied   a  long time ago, even including 
spin~[38-40].

 Although the structure of a  single electromagnetic plane wave is well known,  here we insist on  its  manifestly covariant formulation  as follows. 

\noi The wave vector is a four-vector $k$, null and  oriented toward the future;   the electromagnetic field is a  tensor  
  \beq    F  =   a  (k\wedge  u)  \sin  ( k \cdot x  + \alp ),      \label{def1F}    \eeq
  $u$ is a  constant  spacelike unit vector ($u \cdot u = -1$), $a$  and  $\alp$ scalar constants.
Although $k$ is given, the factor $u$ in the bi-vector    $k\wedge  u$ is not unique since we can add to  $a u^\alp$  an arbitrary  null vector proportional to $k.   \      $
A  vector-potential  for this field  
 is
\beq A^\mu = a u^\mu \cos (k\cdot x + \alp ) ,   \qquad  \qquad      
 k \cdot k  =0
\label{planwav}  \eeq
  The Lorenz gauge condition requires  $u \cdot  k  =0$

\noi  The linear space of four-vectors  can be written  as     
$$  E   =   \cale _{03}     \oplus    \cale _{12}   ,   $$
where  $k  \in  \cale _{03}  $   and    $u   \in    \cale _{12}     $. 
 Note however that this direct-sum decomposition {\em  is not  intrinsically} defined.

\noi  We can use an  orthonormal basis  $(E_\alp )$ defined  such that
$E_0$ points toward the future,     $k =  \ome (E_0  +  E_3 )$  and   $E_2 =u$. 

\noi So we have
\beq   k^ \mu =  ( \ome , 0, 0,  \ome ) \   ,   \qquad  \qquad
 \ome  > 0                  \eeq 
equivalently     $ \     k_ \mu =  ( \ome , 0, 0,  - \ome ) $. Note that  $k \cdot x = \ome  (x_0  + x_3 )  =   \ome  (x^0  -  x ^3 )  $.

\noi  Moreover      $  u^\mu = (0, 0, 1, 0 ) $.

\bigskip
\noi {\sl First integrals}

\noi
The {\sl constants of the motion} are induced by the  symmetries of $K$  in   (\ref{defG}).

\noi   Since $k \cdot x $ does not depend on $x^1 , x^2$ it follows that  $A^\mu$  is invariant under  translation   and  rotation  in $\cale _{12}$,     
 in other words  $A^\mu$   commutes with  $p_1 , p_2 ,$ and  $    j_{12} $.  Further  observe that  $u$, thus also $A$, lies in   $\cale _{12}$ thus
$A \cdot p  +  p \cdot A$ depends only on $x^0 ,  x^3 , p_1, p_2$, which entails  $[G, p_1 ] =  [G, p_2 ] =   0$, therefore    $[K, p_1 ] =  [K, p_2 ] =   0$, so both
$G$ and $K$ are (at least simply)  invariant by   translation~\footnote{In contrast    $G $       fails  to be     invariant  by      {\em rotation}    in    
  $\cale _{12}$,  as  can be seen  by  a direct computation.}
 in $\cale _{12}$ .

\noi  Hence $p_1$ and $p_2$ as {\em first integrals}.

\noi Moreover  $[k \cdot x ,     k\cdot p] = 0$, thus  $[A,  k\cdot p] $ vanishes as well, 
 implying that  $ \    A \cdot p , \quad  p  \cdot A$    and  $ A \cdot A \   $ separately all commute with  $k \cdot p $. Hence $ [G, k \cdot p ]=  [K,   k \cdot p ]  =  0 $, so finally
   $ \       k \cdot p $ {\em  is another constant of the motion}.

\noi In any adapted frame, 
 $\    k \cdot p =  \ome (p_0 +  p_3 )  \       $
 so conservation of  $k \cdot p $  amounts to     having    $p_0 +  p_3  =  {\rm const.}$    
(note that in contrast $(p_0 -  p_3 )$ {\em is not} a constant of the motion).
To summarize: the system enjoys {\em  translation} invariance along $E_1 , E_2 $ and $k$,
more precisely  $K$ as well as $G$ are invariant under these translations.  We could state equivalently: 

 \beprop    $\  K$ and $ G \   $ are  {\em  at least simply}   invariant by translation  along direction $w$  iff  
$\   w \cdot  k  =0$.
\enprop   

\noi This condition characterizes the 3-plane tangent to the light cone along $k$, say  $\Pi _3$.

\bigskip
But  we are interested  in  the possibility of having $\   G \   $   {\em strongly}  translation invariant;    
 the question is as to know  whether $G$ admits directions of {\em strong} translation invariance. 
Translation invariance along $w$ will be {\em strong}  iff additionally    $[G, w \cdot x ] $ vanishes.
One finds that   (irrespective of  the analytical shape of $A$) 
 $   [w \cdot  x ,  G ]  $    vanishes       iff          $w \cdot  u  = 0  $. Finally  

\noi      $G$ {\em is {\em  strongly} translation invariant along $w$ provided}
$$  w \cdot k  =   w  \cdot u  = 0  $$    
In other words $w$  must belong to the two-dimensional vector space $\Pi _2$ spanned by $k$ and $u$.
Unfortunately  it is trivial to check that  in  $\Pi _2$ any direction orthogonal to $u$ is necessarily colinear with  $k$. Thus   
  this vector space {\em fails to admit} any orthonormal frame; it does not provide a unique and straightforward  definition of  
longitudinal and transverse directions. This drawback  with  the single plane wave was pointed out in a previous work; 
in the same paper we already advocated the interest of  considering  rather a superposition of two plane waves~\cite{drozFBS}.

\

 \subsection{Superposition of two  plane waves} 
 Now let us consider the case where the electromagnetic field
 is a superposition of two    plane waves travelling along {\em the same} right line, with respective wave  vectors  
 $\   k, \  l   \   $,  non-colinear, both  null  and future oriented, so that their scalar product is strictly positive, say 
$$\   k \cdot  l =  2  \ome ^2$$ 
With an obvious notation  the field  is  $F = F_\onerom  +  F_\tworom   $,  namely
\beq  F  =  a ( k \wedge u ) \sin  ( k \cdot x  + \alp )      +     b ( l \wedge u ) \sin  ( l \cdot x  + \beta )    \label{defF} \eeq  
where     $\   k, \  l   \   $ are constant null vectors,  $a, b, \alp , \beta$ constant scalars;    $u$ is a  constant spacelike  unit vector ( $\  u\cdot u = -1 \   $) orthogonal to both  $k$ and $l$.

\beprop   Since $k , l$  are given  null and   non-colinear,  the  factor  $u$  in  the bi-vectors  $\     k \wedge u  , \       l \wedge u  \     $ is unique  
\enprop

\noi  Proof  $\    $Looking for a possible  $u'$ such that    $\       k \wedge u'  =  k \wedge u$ and  $  \      l \wedge u'  =  l \wedge u \      $ entails
$u' - u  = \lam k , \qquad    u' - u  = \mu l,  \     $ for some $\lam , \mu  \in \batonR $,  thus     $\lam k = \mu l$  which is impossible unless $\lam =0 =\mu$,         
therefore    $u' = u$. []

\medskip
The most simple   vector-potential,  such that    $F = \dron \wedge  A $,     can be written as 
\beq   A^\mu = A^\mu _\onerom    +  A^\mu   _\tworom    =    W   u^\mu  
\label{defAmu}     \eeq
 with   
\beq   W  =  W_\onerom  +  W_\tworom     =    a  \cos (k \cdot x   +  \alp ) + b \cos (l \cdot x   +  \beta )   \label{defW}  \eeq 
 The   Lorenz gauge condition is ensured by  requiring that   $u \cdot  k  =  u \cdot l  =0$.        


\medskip
\beprop
  Given two {\em  non-colinear}  null vectors,  $\   k, \  l   \   $ future oriented, it is always possible to find  orthonormal  basis   where   $k^0 = l^0$. 
\enprop

\noi Proof $\    $ Define  
$$ E_0 = { (k +l) \over 2 \ome} , \qquad \      
   E_3 = { (k-l ) \over 2 \ome}                       $$
We get 
$$     E_0 ^ 2 =  1 , \qquad   E_3 ^ 2 = - 1,  \qquad  E_0 \cdot E _3 =0     $$  
and $  (k +l) (k -l) =0$, moreover 
 $$ k \cdot E_0   = {k \cdot l \over 2 \ome}  =   \ome ,     \qquad \ 
 k \cdot E_3   =  - {k \cdot l \over 2 \ome}  = -   \ome       $$ 
$$l \cdot E_0 =  \ome , \qquad    l \cdot E_3  =   \ome $$
 and
$(k+l) ^2 =  4 \ome ^2  ,    \qquad \    (k-l ) ^2 = - 4 \ome ^2  $.

\noi So the two-dimensional vector space spanned by $k, l$ has signature $+ -$ and admits $E_0 , E_3 $ as orthonormal basis.

\noi Its orthocomplement in the space ${E}$ of four-vectors,   say  $\cale _{12}$, 
 has elliptic signature; the couple $E_0 , E_3$ can be completed by  $E_1 ,E_2 $ as to form an orthonormal basis, 
    In contrast   to the case of a single plane wave,
in the present case    {\em  the  splitting   $E  = \cale _{03}  \oplus  \cale _{12}$ is   intrinsically defined};   note that $u \in \cale _{12}$. 
[] 

\medskip
\noi   Any basis constructed that way (which  gives to both waves  the same frequency)  will be called a {\sl  monochromatic basis}.
In such a basis we can write
\beq   k^\mu =  ( \ome , 0, 0,  \ome )  \qquad  \quad  
   l^ \mu =  ( \ome , 0, 0,  - \ome )       \eeq

   \subsubsection{One-body motion,  first integrals}

\noi   $A$  depends  only on $x^0 , x^3$, thus  $ [A , p_1 ]  =   [A , p_2 ]   = 0$.  
Moreover         (\ref{defAmu})     implies that $A$ lies in  $\cale _{12}$. It follows that     
$  A \cdot p  +  p \cdot A$  depends only on the mutually  commuting arguments  $x^0 , x^3 , p_1 , p_2 $.  
Finally  $G$ and also $K$ commutes with both $p_1$ and  $ p_2$, say
\beq   [p_1 , G ]      =  [p_1 , K ] =  [p_2 , G ]   =      [p_2 , K ] = 0,              \label{pK}         \eeq
this property of   {\em  simple translation invariance} in the plane
 $\cale _{12}  $  entails  that  {\em   $p_1$ and $p_2$ are constants of the motion}.

 In contrast  $ A^\mu _\onerom$ commutes with $ p_0  + p_3 $ whereas  
              $A^\mu _\tworom$ commutes with $ p_0  - p_3 $,      
moreover      $   [ A^\mu _\onerom  ,  p_0 -  p_3 ]  $ is a function of    $  x^0 -  x^3$  only   whereas    
$   [ A^\mu _\tworom  ,  p_0  +  p_3 ]  $ is a function of    $  x^0   +  x^3$  only.
The most general direction of  ${\cale}_{03}$  can be written as
  $w  =    {w_+} (E_0   +    E_3 )  +    {w_-} (E_0  -  E_3 )   $ ( with  $ w_\pm  $ constant scalars).
One finds   that   
$[A^\mu , w \cdot p ]$ is a sum of two independent functions; it cannot be identically zero,  thus  no direction of   $\cale_{03}$ 
 could generate a  translation leaving $A^\mu$   invariant (in fact   $   A^\mu _\onerom   + A^\mu _\tworom  $ exhibits
{\em no  invariance} at all  in the plane  $\cale_{03}$). 
Similar argument  holds for   $ [A \doot p +  p \doot A , \,  w  \cdot p ]  $  and       $ [ A \doot A , \, w  \cdot p ]   $,  and  finally
 $G$  cannot be translation invariant along a direction of  $\cale_{03}$.

\subsubsection{Strong translation invariance}
Let us prove more briefly   this  statement   announced  several  years ago with a complicated justification~\cite{mpulcian}.

\beprop
The interaction term $G$ corresponding to (\ref{defAmu}) is  strongly translation  invariant along  a unique  direction  $w$, 
defined as  orthogonal to $k,  l$ and  $u$.
\enprop

\noi Proof
 $\qquad  $   in the previous subsection we saw that  $G$ is  (at least simply)  translation invariant along any direction of the plane $\cale _{12}$,
and  by  no  direction of the plane $\cale _{03}$.

\noi 
Thus any possible direction of {\em strong} invariance of  $G$ must be searched only within  the plane $\cale _{12}$. 
 Remind that $u$ being  orthogonal to $k$ and $l$, it belongs to the plane $ \cale _{12}$.

\noi {\em Hereafter we shall   specify further the monochromatic basis by taking 
$E_2  =u$, so we have   $\   u ^\mu =  (0, 0, 1, 0)$}.
This choice determines  { \em the  adapted } monochromatic basis; as a result we can write    the second formula   (\ref{defG})  
  on this form
\beq   G  =    - \half  e  (W p_2  +  p_2  W  )   -  \half e^2     W^2     \label{redefG}   \eeq                   
where the only momentum involved is the component $p_2$.

   Let us now look   for a direction $w$ of {\em strong} translation invariance. 
 In addition to ordinary invariance just characterized above,  we must have that 
$[G ,  w \cdot x ]$ vanishes.   Since  the quadratic piece (with respect to $eA$) of $G$ trivially   commutes with $x$,  we are left with
$$   2 [G ,  w \cdot x ]  =  -e  A^\alp  [ p_\alp ,   w \cdot x ]  -   e   [p_\alp , w \cdot x ] A^\alp         $$   
but        $\    [ p_\alp ,   w \cdot x ]     =  -i  w_\alp$,   therefore        
$[G ,  w \cdot x ]$ vanishes   iff   $w$ is orthogonal to  $A$,  which means  orthogonal to  $u$. 
In the $\cale _{12}$ plane the only direction orthogonal to $u$ is that of $E_1$. []

\medskip
We shall normalize  $w$  by choosing   $w =E_1$,  
 say   $$ w ^\mu  =  (0, 1, 0, 0)                 $$  
This spacelike direction determines a $1  \oplus 3$ decomposition  longitudinal/transverse  in  the linear  space of four-vectors~\footnote{
space  $ \oplus $  three-dimensional hyperbolic,  not to be confused with the usual time $\oplus $ space decomposition.}.

 In agreement with the convention made in §2.2,  the ray spanned by
$w$ will be called  {\sl longitudinal}, whereas the span of $k, l, u$ will be called  the  {\sl transverse} 3-plane.

\medskip
To summarize: in the field  which corresponds to   (\ref{defAmu}) the wave equation is 
$ \disp     \     (p-e A)^2   \psi  =  m^2   \psi               \             $, neither  $k \cdot x$ nor   $l \cdot x$  commutes with the
 squared-mass   operator; in contrast the quantities
$    \quad    p_1 , \quad  \     p_2   \quad         $
are constants of the motion;
they  can be diagonalized,   say
$  \       p_1 \psi  = \rho _1     \psi , \qquad  \         p_2 \psi  = \rho_2      \psi  \        $, with   $    \rho_1 ,    \rho_2$   numerical constants.
 So we can write, up  to a normalization factor  
$$ \psi  =
          {\rm     e}^{ i (\rho_1   x^1   +    \rho_2   x^2  )  }  \     \gam (x^0 ,  x^3 )           $$

In the sequel we shall tackle  the two-body problem, in order to  construct a pair of compatible mass-shell constraints.


   \section{Two-body system in a monochromatic superposition}

\bigskip
Let us  resume our analysis of the two-body problem initiated in subsection 2.2.  As we saw up there,
 in the absence of mutual interaction the motion of each particle would be ruled     by the Hamiltonian generator
$K_a = \half  p_a ^2  +  G_a$.    Now we focus on the situation  characterized by the form
 (\ref{defAmu}) of  electromagnetic vector-potential.  We   use  the adapted frame described in the previous section and extend formula
 (\ref{redefG}) to the two-body system  by  replacing   $x, p, e$  as indicated  in   subsection 2.2; we find
\beq   2G _a =  -e_a  ( W_a  p_{(a) 2}   +     p_{(a) 2}   W_a )   - (e_a  W_a  )^2       \label{2Ga}   \eeq      
where  
\beq   W_a  = a  \cos (k \cdot  q_a   +  \alp ) + b \cos (l \cdot q_a   +  \beta )   \label{defWa}  \eeq
and with the following 

\noi   {\sl  Notation}  $\     $
When there is a risk that particle label  be confused with coordinate label, the former is put 
between parenthesis; no parenthesis  otherwise: for instance    $p_{(1) 2}$ is the second component of the momentum of particle 1. 

\medskip
\noi   {\em Remark  $      $} Strong translation invariance of $G_a$ is a consequence of  that of  $G$,  which stems  from the shape of the 
electromagnetic field as  described in  section 3.2  and stated in Proposition~5,  without any further condition.

\medskip
\noi   Note that each $W_a$ depends only on   $Q^0 , Q^3, z^0, z^3$  while  $G_a$  additionally depends on 
$P_2$  and  $y_2$, through the identities 
\beq  p_{ (1) 2} = \half P_2    + y_2 ,  \qquad  \quad          p_{ (2) 2} = \half P_2    - y_2            \eeq
\noi   \medskip
\noi   It stems from (\ref{pK})  that 
  \beq     [p_{(1) 1} , K_1  ]  =   [p_{(1) 2} , K_1  ]  =   0      \eeq 
 \beq     [p_{(2) 1} , K_2  ]  =   [p_{(2) 2} , K_2  ]  =  0      \eeq 
and it is trivial that 
 \beq     [p_{(1) \alp} , K_2  ]  = [p_{(2) \alp} , K_1  ]  =  0    \eeq
whence we deduce
\beq      [P_1 , K_a ] =  [P_2 , K_a ] =  0       \label{PK} \eeq


The unique longitudinal direction is  $E_1$ with contravariant components 
$ (0, 1, 0, 0, )$ and   now we  have    for any  four-vector
\beq     \xi_\lll  =   -   (\xi  \cdot   w ) \    w ,  \qquad  \quad     
   \xi _ \ttt =  \xi  +  (\xi  \cdot   w ) \    w
\eeq
but  
$  \xi  \cdot   w = \xi  \cdot   E_1 =  \xi _1  = -\xi ^1$.  

\noi
Here  $w$ is spacelike; in any adapted frame we can write for any  couple $\xi , \eta$
\beq  \xi_\lll  \cdot   \eta_\lll=  -   (\xi  \cdot   w )  (\eta  \cdot   w )  =   - \xi _1  \eta _1      \eeq
%

\noi   thus we have  
$$z_\lll  \cdot  P_\lll  =  - z_1   P_1 , \qquad  \       P_\lll ^2  =  -   P_1 ^2 ,       \qquad  \       z_\lll ^2  =  -   z_1 ^2      $$
\beq    {  (z_\lll  \cdot  P )^2  \over   P_\lll ^2  }   =   -   z_1 ^2           \label{cancellor}            \eeq
Owing to this last formula, the third term in the r.h.s. of  (\ref{7IJTP2000})  vanishes and we remain with
\beq             \Zhat =        z_\ttt \cdot z_\ttt \        P ^2    -    ( z_\ttt  \cdot  P_\ttt )^2        \label{bijZhat}    \eeq
analogous with Bijtebier's  formula (see (4.3)(4.4) in \cite{bij89})  concerning the case where the external field was  stationary; but  here the external potential is  strongly  invariant along a {\em spacelike} direction. 

On the other hand, equation   (\ref{varsomKprim}) gets simplified as follows
\beq   K'_1 +K'_2  =  P^2 /  4 + y^2   + G_1  +  G_2    - 2T {y_1   \over   P_1}  -  {T^2  \over  P_1 ^2}     \label{simpsomKprim}    \eeq

\medskip
\noi  Linear momentum is  
$ P _\alp = p_{(1)  \alp} + p_{(2)  \alp}  $.   As we mentioned in subsection 2.2  its longitudinal piece 
$$P_{\lll  \alp} =        - P_1 \    w_ \alp  $$ 
 is conserved, therefore we can diagonalize   $P_1$,   and  fix  its  eigenvalue  say  $\lam_1 \not=  0$ (according to the restriction made in subsection~2.2).
 So we get rid of one spacelike degree of freedom, namely  $Q^1$.

 Moreover, in view of  theorem~2   of \cite{IJTP2009},  we  expect  that   $P_2$ also  is to be conserved.
Let us directly check this point.
 In view of (\ref{PK}) all we have to prove  is that $P_2$ also commutes with $V$, or equivalently that $P' _2$  commutes with $V'$.
But we first observe that
\beprop  $P_2 $ is not affected by transformation (\ref{transbij}) , say  $P'_2 =  P_2$   \enprop
Proof $\    $ $P_2$ is a purely transverse quantity, thus it commutes with $L$.
Then a glance at  (\ref{vardefT})  and   (\ref{PK})    ensures  that $P_2$ commutes also  with $T$, so finally with $LT$ which is the generator of  
the transformation.[]

\noi  Then looking  at    (\ref{ansatz}) the    question is  whether $ P_2$   commutes with   $\Zhat$, which is obvious in  (\ref{7IJTP2000}), so

\noi   $P_2$ {\em  is a constant of the motion}, as expected;  we assign to it a sharp value, say $\lam_2$.

\medskip

We can summarize:
 the survivors of  the Poincar\'e   Lie algebra are $P_1, P_2$, they define a conserved  vector, let it be noted as
\beq   P_{\perp  \alp}       =   (0, P_1 , P_2 , 0 )              \eeq
and the principle of isometric invariance is  satisfied.
Notice  that     $ P'_{\perp } $   is not affected   by transformation  (\ref{transbij}), say              $ P'_{\perp  \alp}  =  P_{\perp  \alp}  $.

\medskip
In contrast to the case without  external field, $P^2$ is not anymore a first integral  whereas    $P_{\perp }^2$ remains conserved.

\noi At this stage it is convenient to remind that  in the absence of external field the coupled wave equations are usually reduced to  a spectral problem for the quantity
$\disp  N =  H_1 + H_2 -  {(H_1 - H_2 )^2   /  P^2}   -  {P^2  /   4}    $  which is intimately related to the properties of  relative motion~\footnote{
The eigenvalue of $-N$ appears denoted as $b^2$  in the  work of Todorov~[18-21];
  divided by the reduced mass it is proportional to the leading term in the development of the mass defect $M-(m_1 +m_2)$, insofar as an {\em isolated system}  is oncerned. 
}.

\noi   In the present case $N$  is not anymore a first integral,  but  now it is natural to consider instead of it    this invariant combination
 \beq      N _{\perp} = H_1 + H_2 -  {   (H_1 - H_2 )^2  \over P_{\perp }^2    }   -   { 1  \over   4   }   P_{\perp } ^2      \label{defNperp}   \eeq

The system  (\ref{somprim})(\ref{difprim}) is to be solved in the external-field representation; we can impose that  $\Psi '$ be an eigenfunction of  $P _{\perp \alp}$ , say
$$  P _{\perp \alp} \    \Psi '   =   I_\alp \      \Psi ' ,   \qquad    I_\alp =     (0,  \lam_1 ,   \lam_2 , 0 )  ,     $$
    this choice  renders   $N'_\perp$ diagonal.  
     The cut-off introduced in Section~2.2   is simply  expressed as       $\lam _1^2    \geq   \varepsilon$,  hence  $I^2  =  - (\lam_1 ^2  +   \lam_2 ^2 )  < 0$.

\medskip
\noi We have 
$$    P^2     \Psi '   =   ( P_0  ^2  -   P_ 3 ^2)    \Psi ' - (\lam _1 ^2  +  \lam _2 ^ 2 )   \Psi '  $$
note that  $P^2 \not=   {P'}^2$.

\subsection{Reducing the wave equations.}
Here we aim at solving the coupled wave equations    (\ref{somprim})(\ref{difprim})  by an eigenstate of  $P_1 , P_2$, say
\beq     \Psi '= {\rm  e}^ {i ( \lam_1  Q^1 + \lam_2  Q^2 ) }  \       \phi (Q ^0 , Q^3 ,   z^\alp )      \eeq
 Let us consider first 
(\ref{difprim})   and  remember  that     $y_1  =  -i  {\dron /  \dron  z^ 1}$.     Since     $y_\lll \cdot P_\lll =- y_1  P_1$      we must have   $y_1  P_1  \    \Psi '  =  - \nu \Psi '$,   but   $ P_1     \Psi '  =  \lam_ 1   \Psi'$,
so   $y_1$ is constant of the motion with eigenvalue  $-  \nu     /  \lam  _1$,   and  (\ref{difprim})   is  to be solved by writting
\beq     \Psi '=   {\rm exp}    i( \lam_1  Q^1 + \lam_2  Q^2    -    {\nu \over \lam _1}  z^1 ) \        \chi (Q ^0 , Q^3 ,   z_\ttt )
\label{soludif}\eeq  
In order to determine   $\chi$ we now develop  equation (\ref{somprim}).

\noi We  separate the coordinates $Q^1 , Q^2 , z^1$ from    $Q^0 , Q^3 , z_\ttt$.
It is clear from    (\ref{soludif})  that  $\Psi'$ is eigenstate not only of  $P_1, P_2$,  but also of $ y_1$,  with respective eigenvalues  
  $\lam _1 , \lam_2$  and  $  - {\nu  / \lam_1}$.              In view of this remark  it is convenient to introduce the  following

 {\sl Notation}: $ \     $   to  any  dynamical variable $\calf (Q,P,z,y ) $  we associate the substitution 
\beq  
 {\underline  \calf}  =  {\rm subs}.    (  \calf   \    |  \     P_1 =  \lam_1 , \    P_2 =  \lam_2 , \     y_1 =    - {\nu   /  \lam_1}     )         \eeq   
so   ${\underline  \calf} $  and    $\calf$  yield   the same result  when applied to  $\Psi '$. 
For instance        
\beq      \soulP ^2     =     P_0^2  -   P^2 _3   -\lam _1^2  -\lam_2^2  ,     \qquad  \qquad     { \soulP} _\lll ^2     =     -  \lam_1^2       
\label{soulPsq}   \eeq  
From  (\ref{2Ga})  we obtain
    \beq  2 \soulG _a  =  - e_a  (W_a   {\underline p}_{ (a) 2}   +   {\underline p}_{ (a) 2}    W_a )  - (e_a  W_a )^2 
\label{2soulGa}    \eeq
where
 \beq   {\underline p}_{ (1) 2} = \half  \lam_2   +  y_2    ,       \qquad  \       {\underline p}_{ (2) 2} = \half  \lam_2   -  y_2                 \eeq
Formula    (\ref{simpsomKprim})   yields   
 \beq    \soulK'_1 +  \soulK'_2 =   {1 \over  4} \soulP^2 - ( {\nu \over \lam_1 }    )^2   +  y_\ttt ^2  +  \soulG_1  +  \soulG_2   +
{1 \over \lam_1^2 } (2 \nu  \soulT  -  \soulT^2 )            \label{somsoulKprima}        \eeq
where   $\soulG _a $   is given  by  (\ref{2soulGa})  above, while  the  expression for  $\soulT$  results from    (\ref{defT}),  that is  
\beq     \soulT  =  y_0 P_0  -  \lam_2  y_2 -P_3  y_3  +  \soulG _1   -   \soulG_2           \eeq

\noi
 But   in order to achieve writting  the explicit form of       (\ref{somprim})  
we still have to evaluate $V' \Psi'$. 

\noi  In view of (\ref{defVzer})(\ref{unipotext})(\ref{ansatz}) 
we consider the action of $\Zhat, P^2 $ and  $y \cdot P$ on the wave function
.  
In the present case $\Zhat$ is given by   (\ref{bijZhat}), hence  
$ \Zhat  \Psi ' =   {\soulZhat} \    \Psi '$  where  of course we define
\beq  \soulZhat  =  z_\ttt \cdot z_\ttt \    {\underline P^2}   -  
                       (z^0 P_0 +  z^3  P_3  +   \lam_2  z^2 )^2
\label{soulZhat}     \eeq
Note that, as  differential operators,   $\soulP ^2$  and   $\soulZhat$  act only on the variables  $Q^0 , Q^3$.
Finally  $V' \Psi' =   \soulVprim \Psi' $, defining
\beq  \soulVprim  = f(\soulZhat , {\underline P^2}, \nu )   ,        \label{soulVprim}  \eeq
in this   expression $f$ encodes all information about mutual interaction; it is  a priori given and the details about its arguments
 are formulas   (\ref{soulPsq})   and            (\ref{soulZhat}). 

   The reduced wave equation thus takes on this form
\beq       (  \soulK'_1 +  \soulK'_2   + 2 \soulVprim  )  \chi  =   \mu  \chi        \label{lastreduc}   \eeq
wherein    
    (\ref{somsoulKprima}) and    (\ref{soulVprim}) are to be inserted. 
In spite of its formal aspect,  (\ref{lastreduc}) cannot be considered as an eigenvalue equation for $\mu$, since $m_1 , m_2$ are  parameters fixed from the outset.
In fact this equation is, in a trivial manner, equivalent to an eigenvalue problem for the quantity defined in   (\ref{defNperp}), say
$$N'_{\perp }  \Psi '  = ( H'_1  +  H'_2   -{\nu ^2  \over   I^2 }   -  {  I^2 \over 4  }       )                          \Psi  '                $$ 
Indeed according to  (\ref{unipotext}) we recall that   $H'_1  +  H'_2  =   K'_1  +  K'_2  + 2 V'$  therefore the  number 
  $ \disp  \sig  = \mu  -   ({\nu ^2  \over I^2 }  +  {I^2  \over 4}) $  is the eigenvalue of $N'_{\perp}$.

\section{Conclusion and outlook}  
From the start    we have discarded the apparently simpler model  involving   a single plane electromagnetic wave, which   actually   is problematic   for our purpose,
 since  it leads to a degenerate case   of strong translation invariance.

In this work we  obtained a pair of compatible mass-shell constraints describing  the motion of two charged  spinless  particles 
subject to a laser made of two counter-propagating plane waves.
The  form of these  equations is given explicitly, in the external-field representation, through formulas   (\ref{somprim})(\ref{difprim}) with help of  
  (\ref{defT})(\ref{2Ga})(\ref{simpsomKprim})  and   (\ref{ansatz}),
  assuming   that the  term of mutual interaction was  known in closed form in the absence of external field.

\noi   The monochromatic superposition of two plane waves (although  it preserves  less symmetries than a single plane wave)  provides  a {\em normal} case of  strong translation invariance.
Moreover (in contrast to the single wave) this superposition allows  us to distinguish, in an intrinsic way,  a preferred   frame of reference (the adapted basis) which could be viewed, rather naturally,  as 
the  {\em  laboratory frame}.

\noi   Enough symmetry of translation  is preserved anyway,  as   to  furnish  two constants of the motion, $P_1, P_2$,  the former associated with strong translation  invariance, and both of them  
in agreement with the principle of  isometric invariance.

\noi On the one hand these first integrals permit to factorize out two degrees of freedom, namely  $Q_1 , Q_2$. On the other hand (\ref{difprim}) leads to the elimination of  $z^1$
so   we remain  with a  unique  reduced wave equation to be solved for 
$\chi (Q_0 , Q_3 ,  z_\ttt )$.  
 We are left with a problem of five degrees of freedom.

Note that, the longitudinal direction being spacelike, the  spacelike relative coordinate $z^1$ is   eliminated instead of the so-called "relative time"; a similar situation also occurs in the  simple case where
 the external field is a constant homogeneous electric field~\cite{drozPhysRevA}.  

At the present stage we are  at least provided with a manifestly covariant  formalism which has the correct limits when either of the interactions vanishes and which satisfies  the principle of isometric invariance.
Naturally  it would be interesting to  renew  the contact with BS equation   in the spirit of    \cite{bijbroek}, and to compare the result  with the present approach;    but now  the variation of  external  field {\em  in time}  might  be a serious complication  for this program.   In the meanwhile it is encouraging that   isometric invariance which was neither  explicitly  required  nor even invoked in the  early  foundations 
of our method~[30,31],
      turns out to be satisfied after all.

Some attention is still required in order to clarify the physical  meaning of  $N_\perp$, but this issue is  already  transparent in the equal-mass case ($\nu = 0$), where $N_\perp$ is just the conserved piece of $N$.

\noi Further work could be devoted to solve the reduced wave equation for a relativistic harmonic oscillator as a  toy model, choosing  $  f =  {\rm const. }\         (P^2 )^{- 1/2}  Z $ in formula~(\ref{defVzer}).

\noi Another open problem in the hope of realistic applications is of course the introduction of spin.


\begin{thebibliography}{3}



\bibitem{audr}  K-P. Marzlin, J. Audretsch,   
Phys.Rev. A {\bf 53}, 1004 (1996)  

\bibitem{bor83}  Ch. J. Bord\'e  {\em in  Advances in Laser Spectroscopy,} Eds.  F.T.  Arecchi, F. Strumia and H. Walther,
Plenum Publishing Corp. (1983)  
\bibitem{IshikBor}           J. Ishikawa,  F. Riehle,  J. Helmcke, Ch. J. Bord\'e, 
Phys.Rev. A    {\bf 49}, 4794-4825 (1994)


\bibitem{ruf}   M. Ruf, G. Mocken, C. Mueller, K. Hatsagortsyan, Ch. Keitel,  
Phys.Rev. Lett  {\bf 102}, 080402 (2009)



\bibitem{heinz}  Th. Heinzl, A. Ilderton, M. Marklund  
 Phys.Lett.  B,  {\bf 692},        250-256 (2010)

\bibitem{barcelo}
An introductive (non-exhaustive) account of modern relativistic dynamics is
 available in
{\it Relativistic Action-at-a-distance, Classical and quantum aspects}, Lecture
Notes in Phys. {\bf 162}, J.Llosa Editor, Springer Verlag (1982) and references
therein.



\bibitem{1droz2}  Ph. Droz-Vincent,  Reports in Math. Phys. {\bf 8}, 79 (1975) 

\bibitem{2droz2}  Ph. Droz-Vincent,   Phys.Rev.D {\bf 19}, 702 (1979).



\bibitem{1bel}    L. Bel, in {\it Differential Geometry and Relativity} p.197, 
 M.Cahen and M.Flato editors,  Reidel Dordrecht (1976) 

\bibitem{2bel}    L. Bel,  Phys. Rev. D {\bf 28}, 1308 (1983) \\


\bibitem{1leutw}  H. Leutwyler and J. Stern, Ann.of Phys.(N-Y){\bf 112}, 94 (1978) 

\bibitem{2leutw}  H. Leutwyler and J. Stern, Phys.Lett.B   {\bf 73}, 75 (1978) 

\bibitem{CratVan}  H.W. Crater and P. Van Alstine,
Phys. Lett. B  {\bf 100}, 166  (1981).

\bibitem{jinr}           I.T. Todorov, JINR Report E2-10125, unpublished (1976).\\

\bibitem{moltod} V.V. Molotkov and I.T. Todorov, Commun. Math. Phys. {\bf 79}, 111 (1981).

 \bibitem{todbarcelo}   
     I.T. Todorov, contribution to reference  \cite{barcelo}.

\bibitem{popov}  V.S. Popov,  Phys. Lett.  A  {\bf 298}, 83 (2002)


\bibitem{tod71}
I.T. Todorov, Phys. Rev. D {\bf 3},2351, (1971)

\bibitem{todZich} I.T. Todorov,  {\em  in}  A. Zichichi ed.{\em Properties of Fundamental Interactions},  vol 9, part C, p. 951;  
  editrice Compositori, Bologna (1973)

\bibitem{riztod}  V. A.  Rizov,  I.T. Todorov,  Sov. J. Part. Nucl. {\bf  6}, 269 (1975).


 \bibitem{todaneva}       V. A.  Rizov,  I.T. Todorov, B. L. Aneva,  Nucl. Phys. B {\bf  98}, 447 (1975). 



 \bibitem{1sazBS}  H. Sazdjian,   Phys.Lett.B {\bf 156}, 381 (1985),

 \bibitem{2sazBS}  H. Sazdjian,   Jour.Math.Phys. {\bf 28},   2618 (1987).



\bibitem{crater}    H.W. Crater, R.L. Becker and C.Y. Wong, P. Van Alstine,
Phys. Rev. D {\bf 46} 5117-5153 (1992).



\bibitem{jallou97} Jallouli and H. Sazdjian,
Ann. Phys.(N-Y),{\bf 253}, 376-426 (1997)



\bibitem{bijbroek}     J. Bijtebier,  J. Broekaert     Nuovo Cim. A,  {\bf   105}, 351 (1992)


\bibitem{1sokol}  S.N. Sokolov,   Theor. Math. Phys. {\bf 36}, 193  (1978)
      
 \bibitem{2sokol}  S.N. Sokolov,   Theor. Math. Phys. {\bf 37}, 1029  (1979)



  \bibitem{IJTP2009}  Ph. Droz-Vincent,
 Int. Journ. Theor. Phys. {\bf 48}, 2177-2189 (2009).



\bibitem{bij89}  J. Bijtebier, Nuovo Cim. A, {\bf 102}, 1285 (1989)

\bibitem{drozNCim}  Ph. Droz-Vincent,
 Nuovo Cimento A {\bf 105 }, 1103-1126 (1992). 
There are several misprints in this paper:  in p.1107 one should read  condition (2.15) instead of condition (3.15).
In Appendices  B and C:    
$Z$ was  skipped from the r.h.s. of formula B.5.
One should read $ K = \overline{H}  +G$  and     
 $\overline{H} _1  + \overline{H} _2  = { P^2 \over 4} + y^2$.  
    

     \bibitem{drozFBS} Ph. Droz-Vincent,    
 Few-Body Systems {\bf 14},  97-115  (1993)    

 \bibitem{mpulcian}    Ph. Droz-Vincent,   Two-body relativistic   system in external field, {\em  in} 
"Constraint Theory and Quantization Methods", Montepulciano 1993,  F. Colomo,  L. Lusanna, G. Marmo Eds. 
World Scientific  (1994).



\bibitem{1nogo}    D.G. Currie, {\em  Journ. Math. Phys. {\bf 4},} 1470 (1963) 
\bibitem{2nogo}    D.G. Currie  {\em Phys.Rev.} {\bf 142}, 817 (1966)
\bibitem{3nogo}    D.G. Currie, T.F. Jordan, E.C.G. Sudarshan, {\em Rev.Mod.Phys.}{\bf 35}, 350 (1963)



\bibitem{IJTP2000}  Ph. Droz-Vincent,
 Int. Journ. Theor. Phys. {\bf 39}, 389-403 (2000).


\bibitem{volkov}      D.M. Volkov,  Zeits. Phys. {\bf  94}    (1935)  250
\bibitem{kibble}  LS.  Brown  and   TWB.  Kibble,   Phys. Rev. {\bf  133A}, 705  (1964)
\bibitem{bergou}  J. Bergou, S. Varr\'o,           J. Phys. A: Math. Gen. {\bf 13},  2823-2837 (1980) 






\bibitem{drozPhysRevA}  Ph. Droz-Vincent,
Phys. Rev. A {\bf 52} , 1837-1844 (1995).








\end{thebibliography}
\end{document}